\documentclass[epj]{svjour}

\usepackage{graphics}

\begin{document}

\title{Properties of pulsed entangled two-photon fields}

\author{Jan Pe\v{r}ina Jr.\inst{1}}
\institute{Joint Laboratory of Optics of
Palack\'{y} University   \\
and Institute of Physics of Academy of Sciences of the Czech Republic \\
\email{perina\underline{\mbox{ }}j@sloup.upol.cz} \\
17. listopadu 50, 772 07 Olomouc \\
Czech Republic }
\date{ }

\abstract{
The dependence of one- and two-photon characteristics of pulsed entangled
two-photon fields
generated in spontaneous parametric down-conversion on the
pump-pulse properties (shape of the pump-pulse spectrum and its
internal structure)
is examined. It is shown that entangled
two-photon fields with defined properties can be generated.
A general relation between the spectra of the down-converted
fields is established. As a special case interference of
two partially overlapping pulsed two-photon fields is studied.
Fourth-order interference pattern of
entangled two-photon fields is investigated
in the polarization analog of the Hong-Ou-Mandel interferometer.}


\maketitle

\section{Introduction}

Over the last decades, the process of spontaneous parametric
down-conversion in nonlinear crystals pumped by
cw lasers has been extensively studied
\cite{MaWo,coinc,Pe}. Nonclassical properties
of entangled two-photon light generated by this process
have been used many times for successful tests of
quantum mechanics \cite{coinc}.

Recently a great deal of attention has been devoted to
the properties of spontaneous parametric down-conversion pumped by
femtosecond pulses. The main reason is that femtosecond pump-pulse duration
may provide time synchronization of several entangled two-photon
fields \cite{three-two,three-two1} and create this way more
than two mutually entangled photon fields. Such fields have already been
successfully used when observing quantum teleportation
\cite{teleport} and generating GHZ states \cite{Zeil,Bouw,Rar}.
The basic theoretical model of down-conversion process
pumped by an (ultrashort) pulse has been developed in
\cite{Ru,Se}. It has been studied in detail for a Gaussian
pump pulse. It has been shown that ultrashort
pump-pulse duration leads in general to a loss
of visibility as a consequence of ``partial''
distinguishability of photons introduced by the pump pulse
\cite{Ru,Se}. Spectral aspects of the process are studied
in \cite{Gr}. Coincidence-count interference pattern
has been already measured in case when the down-converted
fields propagate through narrow frequency filters
\cite{Se,Gr1}. The effect of dispersion in the down-converted
fields on the coincidence-count interference pattern has been studied in
\cite{disper}. Fourth-order interference of two two-photon down-converted
fields has been investigated in the setting where the Mach-Zehnder
interferometer is placed into the path of one of the
output fields coming from the Hong-Ou-Mandel interferometer \cite{Kel}.
Pulsed parametric
frequency conversion can also provide pulsed squeezed light
\cite{Berz}.

In the above mentioned papers a Gaussian pump-pulse spectrum
has been assumed. In the paper we study how the pump-field spectral
characteristics (shape and internal structure of the spectrum)
influence the properties of down-converted fields. This may
find an application in the generation of entangled two-photon
fields with defined properties.
Both one-photon (spectrum and time-dependent mean photon number) and
two-photon (coinci\-dence-count rates) characteristics are investigated.
Two-photon characteristics are considered in the polarization analog
of the Hong-Ou-Mandel interferometer.
Because the down-converted fields are strongly influenced by the
internal structure of the pump spectrum, we consider the case
in which the
overall pump spectrum is composed of the spectra of two mutually coherent
ultrashort pulses. As a special case
two-photon interference of two partially overlapping two-photon fields
is investigated.

\section{Spontaneous parametric down-conversion}

The process of spontaneous parametric down-conversion
is described by the following interaction Hamiltonian \cite{MaWo}:
\begin{equation}    
 \hat{H}_{\rm int}(t) = \int_{-L}^{0} dz \, \chi^{(2)}
 E^{(+)}_p(z,t) \hat{E}^{(-)}_1(z,t) \hat{E}^{(-)}_2(z,t)
 + \mbox{h.c.} ,
\end{equation}
where $ \chi^{(2)} $ is the second-order susceptibility,
$ E^{(+)}_p $ denotes the positive-frequency part of
the electric-field amplitude of the pump field,
and $ \hat{E}^{(-)}_1 $ ($ \hat{E}^{(-)}_2 $)
is the negative-frequency part of the electric-field
operator of down-converted field 1 (2).
The nonlinear crystal extends from $ z=-L $ to $ z=0 $.
The symbol $ \mbox{h.c.} $ means Hermitian conjugate.

The wave function $ |\psi^{(2)}(0,t)\rangle $ describing
an entangled two-photon state at $ z=0 $ (at the output plane
of the crystal)
for times $ t $ sufficiently long so that the nonlinear interaction
is complete can be obtained in the form (for details, see
\cite{disper}):
\begin{eqnarray}      
 |\psi^{(2)}(0,t) \rangle &=&  C_{\psi} \exp[ i(\omega^0_1 +
 \omega^0_2) t ] \int_{-L}^{0} dz \, \int_{-\infty}^{\infty} d\nu_p
  \nonumber \\
 & & \mbox{} \times \int_{-\infty}^{\infty} d\nu_1
  \int_{-\infty}^{\infty} d\nu_2 \,
 {\cal E}_p^{(+)}(0,\nu_p) \hat{a}_1^{\dagger}(\nu_1)
 \hat{a}_2^{\dagger}(\nu_2)
 \nonumber \\
  & & \mbox{} \times
 \exp \left[ i\left(\frac{\nu_p}{v_p} -\frac{\nu_1}{v_1} -
 \frac{\nu_2}{v_2} \right) z  \right]
 \delta( \nu_p - \nu_1 - \nu_2 )  \nonumber \\
 & & \mbox{} \times \exp\left[ i(\nu_1 + \nu_2) t \right]
 |{\rm vac} \rangle ,
\end{eqnarray}
where $ \nu_j = \omega_j - \omega^0_j $ for $ j=p,1,2 $.
The symbol $ {\cal E}_p^{(+)}(0,\nu_p) $ denotes
the spectrum of the positive-frequency
part of the envelope of the pump-field amplitude
at the output plane of the crystal;
$ \hat{a}_1^{\dagger}(\nu_1) $
[$ \hat{a}_2^{\dagger}(\nu_2) $] means the creation
operator of the mode with wave vector $ k_1 $ ($ k_2 $)
and frequency $ \omega^0_1 + \nu_1  $ ($ \omega^0_2 + \nu_2 $)
in down-converted field 1 (2). The symbol $ \omega^0_j $
stands for the central frequency of field $ j $ ($ j=1,2,p $) and
$ 1/v_j $ denotes the inverse of group velocity
of field $ j $ [$ 1/v_j = dk_j/(d\omega_{k_j})
|_{\omega_{k_j}=\omega^0_j} $].
The susceptibility $ \chi^{(2)} $
is included in the constant $ C_{\psi} $.
Frequency- and wave-vector phase
matching for central frequencies ($ \omega^0_p =
\omega^0_1 + \omega^0_2 $) and central wave vectors
($ k^0_p = k^0_1 + k^0_2 $), respectively, are assumed
to be fulfilled when deriving Eq. (2).
We note that spontaneous parametric process is so weak
that the pump-pulse depletion can be omitted even when
intense ultrashort pump pulses are considered.

\section{One-photon characteristics}

The mean number of photons $ {\cal N}_j $ in down-converted
field $ j $ is defined as:
\begin{eqnarray}       
 {\cal N}_j (\tau) &=& \langle \psi^{(2)}(0,t_0) |
 \hat{E}_j^{(-)}(0,t_0 + \tau) \nonumber \\
 & & \mbox{} \times  \hat{E}_j^{(+)}(0,t_0 + \tau)
 | \psi^{(2)}(0,t_0) \rangle ,
\end{eqnarray}
where
\begin{eqnarray}     
 \hat{E}^{(+)}_j(z_j,t_j) &=& \sum_{\nu_j}
 e_j(\nu_j) \hat{a}_j(\nu_j) \nonumber \\
  & & \mbox{} \hskip-4mm \times \exp[ik^v_j(\omega^0_j+\nu_j) z_j -
 i (\omega^0_j + \nu_j ) t_j ] .
\end{eqnarray}
The symbol $ e_j(\nu_j) $ denotes
the amplitude per photon of the mode with the frequency $ \omega^0_j + \nu_j $;
$ k^v_j $ means a wave vector in vacuum in the $ j $th field.
Substituting Eqs. (2) and (4) into Eq. (3) we arrive at the
expression for $ {\cal N}_j $ in the form:
\begin{equation}    
 {\cal N}_j(\tau) = \frac{ (2\pi)^2 |C_{{\cal N}_j}|^2 }{ |D| }
 \int_{-L}^{0} dz \left|  {\cal E}^{(+)}_p(0, \tau -
 D_{pj} z ) \right|^2 ,
\end{equation}
in which
\begin{eqnarray}  
 D_{pj} &=& \frac{1}{v_p} - \frac{1}{v_j} = \Lambda + (-1)^j
 \frac{D}{2}
 , \hspace{1cm} j=1,2, \nonumber \\
 \Lambda &=& \frac{1}{v_p} - \frac{1}{2} \left( \frac{1}{v_1} +
 \frac{1}{v_2} \right) , \nonumber \\
 D &=& \frac{1}{v_1} - \frac{1}{v_2} .
\end{eqnarray}
The symbol $ {\cal E}^{(+)}_p(0,t) $ denotes the positive
frequency part of the envelope of the pump-field
amplitude
at the output plane of the crystal; $ C_{{\cal N}_j} $
is a constant [$ C_{{\cal N}_j} = \sqrt{2\pi}
C_{\psi} $ $ e_j(\nu_j=0) $].
Because photons are in the nonlinear process generated in pairs,
it holds that $ \int_{-\infty}^{\infty} d\tau
{\cal N}_1(\tau) = \int_{-\infty}^{\infty} d\tau
{\cal N}_2(\tau) $.

If we consider the case in which $ D_{pj} L \ll
\tau_{\rm char} $, where $ \tau_{\rm char} $ is a characteristic
time of the change of pump-field intensity, then, according to
Eq.~(5), the time dependence of $ {\cal N}_j(\tau) $ resembles that
of the pump field. This means that one-photon multimode Fock-state fields
with a given mean-photon-number time dependence can be generated
if the pump-field intensity is suitably chosen (see Fig.~2 in
Subsec.~5.1).

The spectrum $ {\cal S}_j $ of down-converted field
$ j $ defined as
\begin{equation}         
 {\cal S}_j(\nu_j)  = \langle \psi^{(2)}(0,t_0) |
  e^*_j(\nu_j) \hat{a}^\dagger_j(\nu_j) e_j(\nu_j)
  \hat{a}_j(\nu_j) | \psi^{(2)}(0,t_0)
 \rangle
\end{equation}
can be obtained in terms of the pump-pulse spectrum:
\begin{eqnarray}      
 {\cal S}_j(\nu_j) &=& |C_{{\cal S}_j}(\nu_j)|^2 \int_{-\infty}^{\infty}
 d\nu_p \left| {\cal E}^{(+)}_p(0,\nu_p) \right|^2  \nonumber \\
 & & \mbox{} \times L^2 {\rm sinc}^2
 \left[ \frac{L}{2} \left( D_{p\,3-j} \nu_p - D\nu_j \right) \right] ,
\end{eqnarray}
where
\begin{equation}      
 {\cal E}_p^{(+)}(0,\nu_p) =  \frac{1}{2\pi} \int_{-\infty}^{\infty}
 dt \, {\cal E}_p^{(+)}(0,t) \exp( i \nu_p t) ,
\end{equation}
$ {\rm sinc}(x) = \sin(x) / x $, and
$ C_{{\cal S}_j}(\nu_j) = C_{\psi} e_j(\nu_j) $.
According to Eq. (8), the spectrum of a down-converted field
is obtained as a convolution of the pump-pulse spectrum with a
function ($ {\rm sinc}^2 $) characterizing phase-matching
of the interacting fields in the nonlinear crystal.
Oscillations in the spectrum $ {\cal S}_j(\nu_j) $
occur for longer crystals if the pump-field spectrum
$ |{\cal E}^{(+)}_p(0,\nu_p)|^2 $ has a peaked structure
(for an example, see Fig.~3 in Subsec.~5.1).
The higher the number of peaks in the pump-field spectrum is the
smaller the amplitude and period of the oscillations are.
For shorter crystals the $ {\rm sinc}^2 $ function
in Eq. (8) is wide and smooths out the peak structure of
the pump-pulse spectrum $ |{\cal E}^{(+)}_p(0,\nu_p)|^2 $. This
results in a suppression of oscillations in the spectrum
$ {\cal S}_j(\nu_j) $.

Eq. (8) can be inverted and
the pump-field spectrum $ \left| {\cal E}^{(+)}_p(0,\nu_p) \right|^2 $ may
be determined from the spectrum $ {\cal S}_j $.
This is useful for the generation of a down-converted
field (in multimode Fock state) with a required spectrum,
because the inverse formula to that in Eq. (8) provides
a suitable profile of the pump-field spectrum.

If $ |D_{p1}| \geq |D_{p2}| $, the spectrum $ {\cal S}_1 $ can be expressed
in terms of the spectrum $ {\cal S}_2 $ as follows
($ |C_{{\cal S}_j}(\nu_j)|^2 \equiv c_{{\cal S}_j} $ is assumed to be
frequency independent for $ j=1,2 $):
\begin{equation}    
 {\cal S}_1(\nu_1) = \frac{c_{{\cal S}_1} }{ c_{{\cal S}_2} }
 \int_{-\infty}^{\infty} d\nu_2
 p_{|D|L,d}(\nu_1 - \nu_2) {\cal S}_2
 \left(-\frac{\nu_2}{d} \right) ,
\end{equation}
where $ d = D_{p2} / D_{p1} $ and
\begin{equation}    
 p_{x,y}(\nu) = \frac{1}{\pi y} \int_{0}^{x} dt \,
 \frac{x-|t|}{x-y|t|} \cos(\nu t) .
\end{equation}
The relation in Eq. (10) between the spectra
of the down-converted fields is rather general and holds for an
arbitrary pump field. This relation is a consequence of
entanglement of photons emerging during their generation.
It shows how photons in modes of the down-converted fields
are correlated. Because the function $ p(\nu) $ has a peak around
$ \nu = 0 $ and goes to zero for larger values of $ \nu $,
the mode with a given $ \nu_1 $
has the strongest correlation with the mode of
frequency $ \nu_2 = - \nu_1 / d $ [see Eq. (10)]. It is worth to note
that for $ d < 0 $ the strongest correlation occurs
between modes at the same sides of the spectra.

\section{Two-photon characteristics}

Two-photon properties of down-converted fields determine results
in coincidence-count measurements. Such measurements
are conveniently described in terms of a two-photon amplitude
$ {\cal A}_{12} $ defined as follows:
\begin{eqnarray}    
  {\cal A}_{12}(\tau_1,\tau_2)
   &=& \langle {\rm vac} |
  \hat{E}^{(+)}_1(0,t_0+\tau_1) \nonumber \\
 & & \mbox{} \times \hat{E}^{(+)}_2(0,t_0+\tau_2)
  |\psi^{(2)} (0,t_0)\rangle .
\end{eqnarray}
Using the state $ |\psi^{(2)} (0,t_0)\rangle $ given in
Eq.~(2), we arrive at the expression ($ \omega^0_1 = \omega^0_2 $
is assumed):
\begin{eqnarray}   
 {\cal A}_{12}(T_0,\tau) &=& C_{\cal A} \frac{1}{|D|}
 \exp(-2i\omega^0_1 T_0) {\rm rect} \left( \frac{\tau}{DL} \right)
  \nonumber \\
 & & \mbox{} \times
 {\cal E}_p^{(+)}\left(0,\frac{\Lambda}{D}\tau + T_0 \right) ;
\end{eqnarray}
$ \tau = \tau_1 - \tau_2 $ and $ T_0 = (\tau_1 + \tau_2)/2 $.
The symbol $ {\rm rect} $ means the rectangular function
($ {\rm rect}(x) = 1 $ for $ 0 < x < 1 $, $ {\rm rect}(x) = 0 $
otherwise); $ C_{\cal A} = 2\pi e_1(\nu_1=0) e_2(\nu_2=0) C_{\psi} $.

We further consider one of typical interferometric
configurations, the polarization analog of the Hong-Ou-Mandel interferometer
(see Fig.~1).
\begin{figure}
 \resizebox{0.45\textwidth}{!}{\includegraphics*{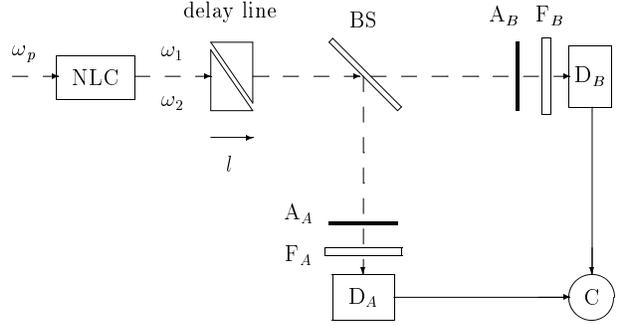}}
 \caption{Sketch of the system for coincidence-count
 measurement:
 pump pulse at the frequency $ \omega_p $ generates
 in the nonlinear crystal NLC down-converted photons at
 the frequencies $ \omega_1 $ and $ \omega_2 $. They
 propagate through a delay line of the length $ l $
 and are detected at the detectors $ {\rm D}_{\rm A} $ and
 $ {\rm D}_{\rm B} $; BS denotes a beamsplitter,
 $ {\rm A}_{\rm A} $ and $ {\rm A}_{\rm B} $ are analyzers,
 $ {\rm F}_{\rm A} $ and $ {\rm F}_{\rm B} $ are frequency filters, and
 C means a coincidence device. }
\end{figure}
Assu\-ming type-II parametric down-conversion,
two mutually perpendicularly polarized photons
occur at the output plane of the crystal.
They propagate through a birefringent
material of a variable length $ l $ and then hit a 50/50~\% beamsplitter.
The coincidence-count rate $ R_c $ is then given by
the number of simultaneously detected photons at both detectors
$ D_{\rm A} $ and $ D_{\rm B} $ in a given time interval.
Analyzers rotated by 45 degrees with respect
to ordinary and extraordinary polarization directions
of the nonlinear
crystal enable quantum interference between
two paths leading to a coincidence count;
either a photon from field 1 is detected
by the detector $ {\rm D}_A $ and a photon from field 2 by
the detector $ {\rm D}_B $ or vice versa.
The normalized coincidence-count rate $ R_n $ in this setting can be
expressed as follows \cite{disper}:
\begin{equation}    
 R_n(l) = 1 - \rho(l) ,
\end{equation}
where
\begin{eqnarray}   
 \rho(l) &=& \frac{1}{2R_0} \int_{-\infty}^{\infty} dt_A \,
 \int_{-\infty}^{\infty} dt_B \, \nonumber \\
 & & \mbox{} \times {\rm Re} \left[  {\cal A}_{12}\left(t_A- \frac{l}{g_1},
 t_B- \frac{l}{g_2} \right) \right. \nonumber \\
 & & \left. \mbox{} \times
 {\cal A}^*_{12} \left(t_B-\frac{l}{g_1},t_A-\frac{l}{g_2} \right) \right]
\end{eqnarray}
and
\begin{equation}       
 R_0 = \frac{1}{2} \int_{-\infty}^{\infty} dt_A \,
 \int_{-\infty}^{\infty} dt_B \,
 \left| {\cal A}_{12}(t_A,t_B) \right|^2 .
\end{equation}
The symbol $ {\rm Re} $ means real part;
$ l $ is the length of a birefringent optical material
(compare Fig. 1) with the group velocities $ g_1 $ and $ g_2 $
appropriate for the down-converted fields 1 and 2, respectively.
Using the expression for $ {\cal A}_{12} $ in Eq. (13),
we get ($ D > 0 $ is assumed):
\begin{eqnarray}   
 \rho(l) &=& \frac{|C_{\cal A}|^2 }{2R_0 D^2}
  \int_{-DL/2 +|\Delta\tau_l|}^{DL/2 -|\Delta\tau_l|} d\tau
  \int_{-\infty}^{\infty} dT_0 \, \nonumber \\
 & & \mbox{} \hskip-4mm \times {\rm Re} \left\{ {\cal E}^{(+)}_p
  \left(0, \frac{\Lambda}{D}\tau + T_0 \right) {\cal E}^{(-)}_p
  \left(0, - \frac{\Lambda}{D}\tau + T_0 \right) \right\} ,
  \nonumber \\
 & &  \\
 R_0 &=& \frac{|C_{\cal A}|^2 }{2 D^2}
  \int_{0}^{DL} d\tau
  \int_{-\infty}^{\infty} dT_0 \, \left|{\cal E}^{(+)}_p
  \left(0, \frac{\Lambda}{D}\tau + T_0 \right) \right|^2 ,
  \nonumber \\
 & &
\end{eqnarray}
where $ \Delta\tau_l = \tau_l - DL/2 $ and
\begin{equation}   
 \tau_l = \left( \frac{1}{g_2} - \frac{1}{g_1} \right) l .
\end{equation}

The analysis of the interference term $ \rho(l) $ in Eq. (17)
for an arbitrary pump field shows, that $ \rho(l) $ is nonzero
only in the interval $ 0 \leq \tau_l \leq DL $.
The interference pattern has the shape of a dip of the
width $ DL $. The change of envelope of the pump field
leads only to a small modification of this shape. On the
other hand internal structure of the pump field may result
in the occurrence of peaks at the bottom of the dip
(for an example, see Fig.~4 in Subsec.~5.2).
This behaviour follows if we consider the general relation between the
shape of the two-photon amplitude and
the profile of the interference pattern (for details, see
\cite{disper}) for the two-photon
amplitude $ {\cal A}_{12}(T_0,\tau) $ given in Eq. (13).

Both one- and two-photon characteristics depend strong\-ly
on the internal structure of the pump-pulse spectrum.
In experiment pump-pulse spectra with interesting internal
structures can be obtained, e.g., if we add spectra of several
mutually coherent ultrashort pulses \cite{Rudolph}.
The pulses can be mutually
delayed and have in general different widths and phase modulations
of their spectra. We further pay attention to the case when the
pump field is composed of two ultrashort pulses.

\section{Effects of the internal structure of the pump-pulse
spectrum}

We assume that the pump field consists of
two mutually delayed pulses with the amplitudes $ {\cal E}_{p1} $
and $ {\cal E}_{p2} $:
\begin{equation}    
 {\cal E}^{(+)}_{p}(0,t) = {\cal E}^{(+)}_{p1}(0,t)
 + \exp(i\phi) {\cal E}^{(+)}_{p2}(0,t+\vartheta) ,
\end{equation}
where $ \vartheta $ denotes a mutual delay between the pulses and
$ \phi $ stands for a relative phase of the pulses.

Pulses with Gaussian envelopes \cite{Rudolph} are used in numerical
calculations:
\begin{equation}  
 {\cal E}^{(+)}_{pj}(0,t) = \xi_{j} \exp \left( -
 \frac{ 1+ ia_j }{ \tau_j^2 } t^2 \right) , \hspace{1cm}
 j=1,2 .
\end{equation}
The symbol $ \xi_j $ denotes the amplitude of the $ j $th pulse
with duration $ \tau_j $ and chirp parameter $ a_j $.
The spectrum $ {\cal E}^{(+)}_{pj}(0,\nu_j) $ determined
according to the definition in Eq. (9) is obtained in the form:
\begin{eqnarray}  
 {\cal E}^{(+)}_{pj}(0,\nu_j) &=& \xi_{j} \frac{ \tau_j }{ 2
 \sqrt{\pi} \sqrt{1 + ia_j} } \exp \left( -
 \frac{ \tau_j^2 }{ 4(1+ia_j) } \nu_j^2 \right) , \nonumber \\
 & & \mbox{}  \hspace{1cm} j=1,2 .
\end{eqnarray}

\subsection{One-photon characteristics}

For short crystals the time dependence of the mean number
of photons $ {\cal N}_j(t) $ in mode $ j $ resembles
that of the pump field (see Sec. 3). So, e.g.,
if the pump field consists of two femtosecond pulses of the same duration
and one has no chirp whereas the other one is highly chirped, the
overall pump field as well as $ {\cal N}_j(t) $ has a peaked structure
(see Fig.~2).
\begin{figure}
 \resizebox{0.45\textwidth}{!}{\includegraphics*{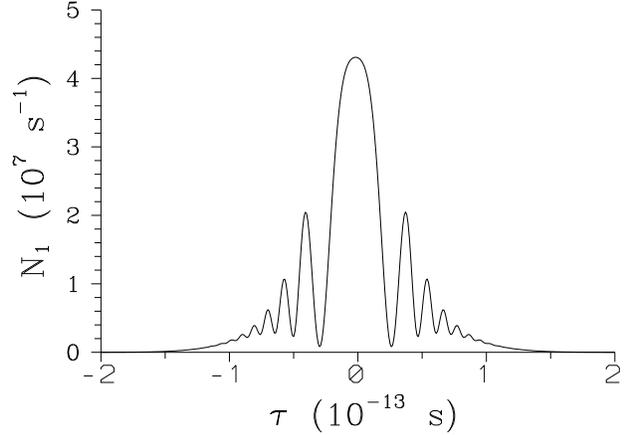}}
 \caption{Mean number of photons  $ N_1(\tau) $ of the signal field;
 $ \tau_1 = 1 \times 10^{-13} $~s, $ \tau_2 = 0.5 \times 10^{-13} $~s,
 $ a_1 = 0 $, $ a_2 = 10 $, $ \xi_1 = \xi_2 = 1 $,
 $ L = 0.05 $~mm, $ \phi = 0 $~rad, $ \theta = 0 $~s,
 $ |C_{{\cal N}_1}|^2 = 1 $~m$ {}^{-2} $.
 In Figs. 2--9,
 values of the inverse group velocities appropriate for the
 BBO crystal \cite{param} with type-II interaction
 at the pump wavelength $ \lambda_p = 413 $ nm
 and at the down-conversion wavelengths
 $ \lambda_1 = \lambda_2 = 826 $ nm apply:
 $ 1/v_p = 56.85 \times 10^{-13} $ s/mm,
 $ 1/v_1 = 56.14 \times 10^{-13} $ s/mm, and
 $ 1/v_2 = 54.30 \times 10^{-13} $ s/mm.
 The optical material for the delay line is assumed
 to be quartz, for
 which $ 1/g_1 = 51.25 \times 10^{-13} $ s/mm and
 $ 1/g_2 = 51.59 \times 10^{-13} $ s/mm.}
\end{figure}

Oscillations occur in one-photon spectra $ S_1 $ and $ S_2 $,
see Fig. 3 for the signal spectrum $ S_1 $.
\begin{figure}
 \resizebox{0.45\textwidth}{!}{\includegraphics*{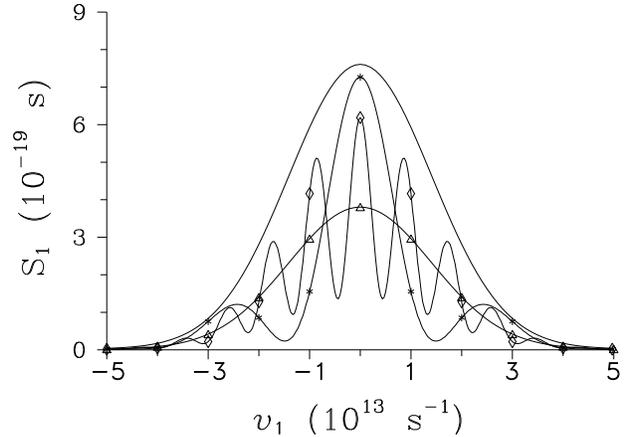}}
 \caption{Spectrum $ S_1(\nu_1) $ of the signal field
 for various values of the delay $ \vartheta $; $ \vartheta = 0 $~s
 (solid curve without symbols), $ \vartheta = 3 \times 10^{-13} $~s
 ($ \ast $), $ \vartheta = 10 \times 10^{-13} $~s ($ \diamond $),
 and $ \vartheta = 50 \times 10^{-13} $~s ($ \triangle $);
 $ \tau_1 = \tau_2 = 1
 \times 10^{-13} $~s, $ a_1 = a_2 = 0 $, $ \xi_1 = \xi_2 = 1 $,
 $ L = 10 $~mm, $ \phi = 0 $~rad, and
 $ |C_{{\cal S}_1}|^2 = 1 $~m$ {}^{-2} $.}
\end{figure}
The period as well as the amplitude of these oscillations
decrease with increasing values of the delay $ \vartheta $
as it has been discussed in Sec.~3.
A nonzero delay $ \vartheta $ causes oscillations in the pump-field
spectrum $ |{\cal E}^{(+)}_p(0,\nu_p)|^2 $ which are transferred
through the phase-matching function [$ {\rm sinc}^2 $ in Eq.
(8)] into the spectra of the down-converted fields. These oscillations are
well pronounced for longer crystals.
If the pump pulses are in phase ($ \phi = 0 $~rad), there is
maximum in the center of the spectrum. There occurs a local minimum
in the center of the spectrum providing that the pump pulses are out
of phase.
Alternatively we may consider the overall down-converted field
to be composed of two contributions from two down-conversion
processes (pumped by two pump pulses) and then the origin of the
oscillations lies in a coherent summation of these contributions
in each frequency mode.

A suitable choice of the pump-field characteristics may provide
down-converted fields with required one-photon properties.
Down-converted fields with given properties can also be obtained
by ``passive methods'', e.g., by frequency filtering of fields
already generated in parametric process. However, entanglement
of photons between two down-converted fields is partially disturbed in
this case.

\subsection{Two-photon characteristics}

The formulas for the quantities $ \rho $ (17) and $ R_0 $ (18)
can be recast into the following form if the pump field given
in Eq. (20) is taken into account:
\begin{eqnarray}   
 \rho(l,\vartheta,\phi) &=& \rho_1(l) + \rho_2(l,\vartheta,\phi) , \\
  \rho_1(l) &=& \frac{ |C_{\cal A}|^2 }{ 2R_0 D^2 }
  \int_{-DL/2 +|\Delta\tau_l|}^{DL/2 -|\Delta\tau_l|} d\tau
  \int_{-\infty}^{\infty} dT_0 \nonumber \\
  & & \mbox{} \times {\rm Re} \left\{ \sum_{j=1,2}
  {\cal E}^{(+)}_{pj} \left(0, \frac{\Lambda}{D} \tau + T_0 \right)
   \right. \nonumber \\
  & & \left. \mbox{} \times
  {\cal E}^{(-)}_{pj} \left(0, - \frac{\Lambda}{D} \tau + T_0
   \right) \right\} , \\
  \rho_2(l,\vartheta,\phi) &=& \frac{ |C_{\cal A}|^2 }{ R_0 D^2 }
  \int_{-DL/2 +|\Delta\tau_l|}^{DL/2 -|\Delta\tau_l|} d\tau
  \int_{-\infty}^{\infty} dT_0 \nonumber \\
  & & \mbox{} \times {\rm Re} \left\{ \exp(-i\phi)
  {\cal E}^{(+)}_{p1} \left(0, \frac{\Lambda}{D} \tau + T_0 \right)
  \right. \nonumber \\
  & & \left. \mbox{} \times
  {\cal E}^{(-)}_{p2} \left(0, - \frac{\Lambda}{D} \tau + T_0 + \vartheta
  \right) \right\}
  , \\
  R_0(\vartheta,\phi) &=& R_{01} + R_{02}(\vartheta,\phi) ,\\
  R_{01} &=& \frac{ |C_{\cal A}|^2 L }{ 2|D| }
  \int_{-\infty}^{\infty} dT_0 \left[ \sum_{j=1,2}
   \left|{\cal E}^{(+)}_{pj}(0,T_0) \right|^2 \right]  , \\
  R_{02}(\vartheta,\phi) &=& \frac{ |C_{\cal A}|^2 L }{ |D| }
  \int_{-\infty}^{\infty} dT_0 \, {\rm Re} \left\{ \exp(-i\phi)
   {\cal E}^{(+)}_{p1}(0,T_0) \right. \nonumber \\
  & & \left. \mbox{} \times {\cal E}^{(-)}_{p2}(0,T_0+\vartheta)
   \right\} .
\end{eqnarray}
The expressions for $ \rho_1 $, $ \rho_2 $, $ R_{01} $,
and $ R_{02} $ for Gaussian pump pulses are contained in Appendix.

We demonstrate the general conclusions given in Sec.~4 considering
two Gaussian pump pulses of different time durations and being out of
phase. The overall pump field then has positive
values at the edges and is negative in the center and so the
corresponding fourth-order interference pattern consists of
three dips (see Fig. 4).
\begin{figure}
 \resizebox{0.45\textwidth}{!}{\includegraphics*{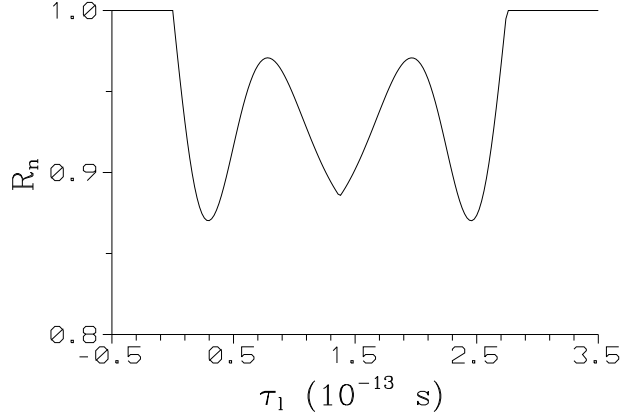}}
 \caption{Interference pattern in the normalized coincidence-count
 rate $ R_n(\tau_l) $ shows three dips;
 $ \tau_1 = 1 \times 10^{-13} $~s, $ \tau_2 = 0.5 \times 10^{-13} $~s,
 $ a_1 = a_2 = 0 $, $ \xi_1 = 1 $, $ \xi_2 = 1.5 $,
 $ L = 1.5 $~mm, $ \phi = \pi $~rad, and
 $ |C_{\cal A}|^2 = 10 $~m$ {}^{-2} $.}
\end{figure}

\subsection{Interference of two entangled two-photon fields}

Physically interesting case occurs if the overall pump field
consists of two partially overlapping ultrashort pulses.
Then the overall down-converted field can be considered as
composed of two partially overlapping two-photon fields
[see the expression for $ {\cal A}_{12} $ in Eq. (13)].
The two two-photon fields are mutually coherent and interfere.
In the following we study the fourth-order interference of these
fields in detail.

Fourth-order interference influences coincidence-count probability.
This probability is linearly proportional to the
quantity $ R_0(\vartheta,\phi) $ which is shown in Fig. 5 as a
function of the mutual delay $ \vartheta $ of the pump pulses in case
when two pump pulses
are in phase ($ \phi = 0 $~rad).
\begin{figure}
 \resizebox{0.45\textwidth}{!}{\includegraphics*{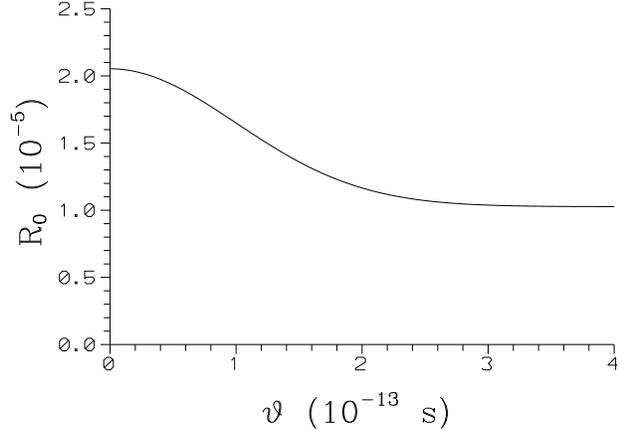}}
 \caption{Probability of a coincidence count $ R_0 $
 as a function of the delay $ \vartheta $; $ \tau_1 = \tau_2 = 1
 \times 10^{-13} $~s, $ a_1 = a_2 = 0 $, $ \xi_1 = \xi_2 = 1 $,
 $ L = 1.5 $~mm, $ \phi = 0 $~rad, and
 $ |C_{\cal A}|^2 = 10 $~m$ {}^{-2} $.}
\end{figure}
The coincidence-count probability
decreases with increasing $ \vartheta $ and is twice for
completely overlapping entangled
two-photon fields ($ \vartheta = 0 $~s) than in the case where there is no
overlap ($ \vartheta \rightarrow \infty $~s). This dependence
reflects constructive interference between two entangled two-photon
fields having its origin in the mutual coherence of the pump pulses.

The fourth-order interference
pattern in the polarization analog of the Hong-Ou-Mandel interferometer
behaves as follows.
The visibility $ V $ of the coincidence-count pattern
(coincidence-count dip \cite{MaWo}) as a function of
the delay $ \vartheta $ is shown in Fig. 6.
\begin{figure}
 \resizebox{0.45\textwidth}{!}{\includegraphics*{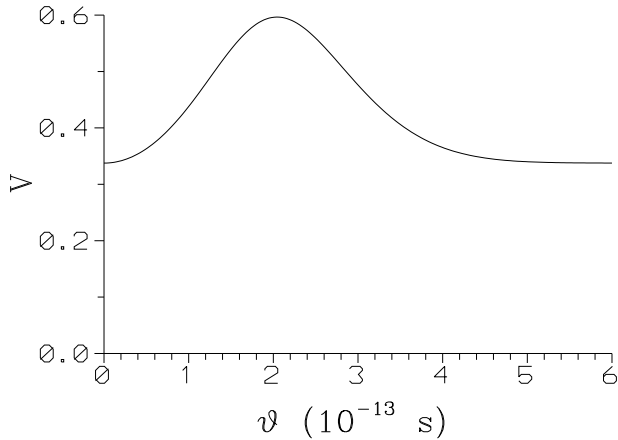}}
 \caption{Visibility $ V $ [$ V = (R_{n,\rm max} - R_{n, \rm
 min}) / (R_{n,\rm max} + R_{n, \rm min}) $] as a function of the
 delay $ \vartheta $; values of the parameters are
 the same as in Fig.~5.}
\end{figure}
For $ \vartheta=0 $~s the entangled
two-photon fields completely overlap, so in fact,
there is only one entangled two-photon field with a given visibility. As the
delay $ \vartheta $ increases, two entangled two-photon
fields are gradually formed at the output plane of the crystal.
When the delay
$ \vartheta $ increases, the overlap of
the two entangled two-photon fields becomes smaller and higher values of
the visibility $ V $ occur.
This means that distinguishability of the signal and idler
photons decreases with increasing $ \vartheta $ because the higher
the visibility the higher
the distinguishability of the signal and idler photons \cite{Se}.
For greater values of $ \vartheta $ the visibility $ V $
decreases.
When the pump pulses are delayed so much that they do not overlap
there are two non-overlapping down-converted fields and
the visibility is back to the value appropriate for
$ \vartheta = 0 $~s. The increase of the visibility $ V $
caused by partially overlapped entangled two-photon fields might
be useful in various multiparticle experiments
for which high visibilities are required.
When the distinguishability of the signal
and idler photons is minimum ($ V $ is maximum), we can already distinguish
two entangled two-photon fields in the signal-field mean photon number
$ {\cal N}_1(t) $. We note
that the value of the delay $ \vartheta $ for which two
entangled two-photon fields in a down-converted beam can be distinguished
differs for the
signal and idler fields as a consequence of different group
velocities in the nonlinear crystal.

The relative phase $ \phi $ influences the shape of
coincidence-count pattern. There might occur a peak at the bottom
of the dip if the pump pulses are not in phase ($ \phi \neq 0
$~rad, see Fig. 7).
\begin{figure}
 \resizebox{0.45\textwidth}{!}{\includegraphics*{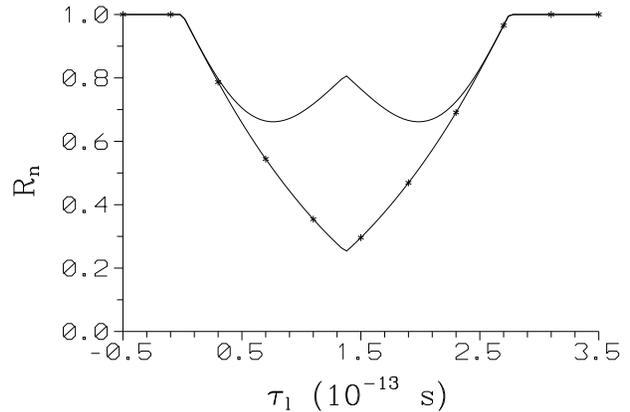}}
 \caption{Interference pattern in the normalized coincidence-count
 rate $ R_n(\tau_l) $ for various phases $ \phi $:
 $ \phi = \pi $~rad (solid curve without symbols)
 and $ \phi = 0 $~rad (solid curve with $ \ast $); $ \vartheta = 2.04
 \times 10^{-13} $~s; values of the other parameters are the same
 as in Fig. 5.}
\end{figure}
This then results in a loss of visibility.
The reason is that if the pump pulses are not in phase
the overall two-photon amplitude $ {\cal A}_{12}(T_0,\tau) $ gets phase
modulation along $ T_0 $ [compare the expression for $ {\cal A}_{12} $
in Eq. (13)] which then results in the loss of visibility
(for a relation between the shapes of the two-photon
amplitude $ {\cal A}_{12} $ and the coincidence-count pattern
described by $ R_n $, see \cite{disper}).
A typical dependence of the visibility $ V $ as a function of
$ \phi $ is depicted in Fig. 8.
\begin{figure}
 \resizebox{0.45\textwidth}{!}{\includegraphics*{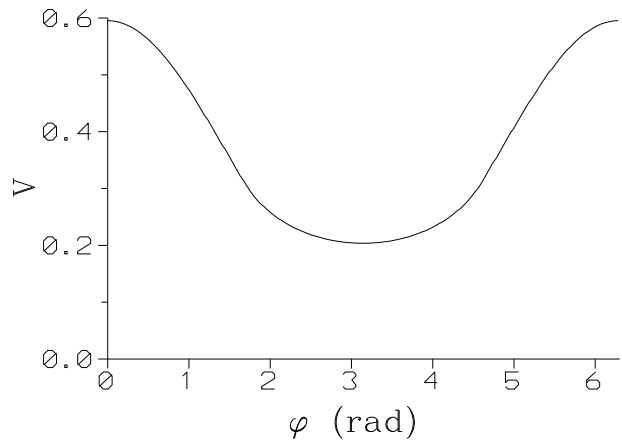}}
 \caption{Visibility $ V $ as a function of the relative
 phase $ \phi $; $ \vartheta = 2.04 \times 10^{-13} $~s,
 values of the other parameters are the same as in Fig. 5.}
\end{figure}

The value of the delay $ \vartheta_{\rm max} $ for which there is a
maximum visibility of coincidence-count pattern increases
with the increasing pump pulse duration $ \tau_0 $ (we assume $ \tau_1 =
\tau_2 = \tau_0 $). This is demonstrated in Fig. 9.
\begin{figure}
 \resizebox{0.45\textwidth}{!}{\includegraphics*{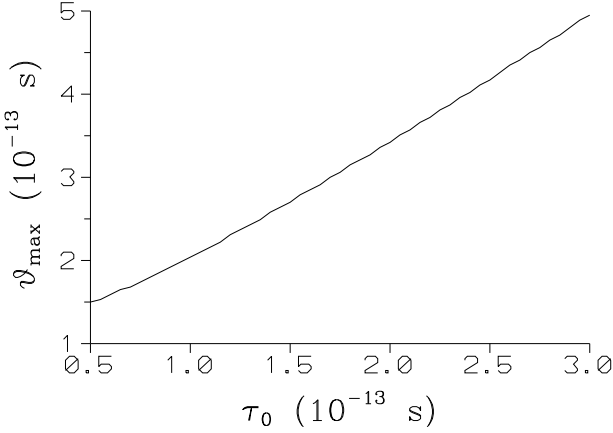}}
 \caption{Delay $ \vartheta_{\rm max} $ corresponding
 to the maximum of the visibility $ V(\vartheta) $ as a function
 of the pump pulse duration $ \tau_0 $
 ($ \tau_0 = \tau_1 = \tau_2 $); $ a_1 = a_2 = 0 $,
 $ \xi_1 = \xi_2 = 1 $, $ L=1.5 $~mm, $ \phi = 0 $~rad,
 and $ |C_{\cal A}|^2 = 10 $~m$ {}^{-2} $.}
\end{figure}
The value
of $ \vartheta_{\rm max} $ is influenced also by pump-pulse chirp. The
higher the chirp the lower the value of $ \vartheta_{\rm max} $.
However, nonzero chirp also results in the loss of visibility
\cite{disper} because it leads to a phase modulation of the two-photon
amplitude $ {\cal A}_{12} $. Using this fact, the influence of chirp
can be eliminated from the measured visibilities and the pump-pulse duration
$ \tau_0 $ may be in principle determined
from the measured value of $ \vartheta_{\rm max} $.

\section{Conclusions}

We have studied both one- and two-photon properties
of entangled two-photon fields generated by spontaneous
parametric down-conversion pumped by fields with arbitrary
characteristics. The spectrum of a down-converted field may be
expressed as a convolution of the pump-field spectrum
with the function originating in phase-matching of waves
in a nonlinear crystal. The spectra of the down-converted
fields are mutually strongly correlated and a general relation between them
independent of the pump-field parameters has been established.
Both one- and two-photon properties depend strongly on the internal
structure of the pump field. For example, there may occur oscillations
in the spectrum of a down-converted field.
One-photon multimode Fock-state fields with defined
characteristics (intensity or spectrum profile) can be generated
if the pump-field parameters are suitably chosen.
For more complex pump fields a typical dip in the
coincidence-count interference pattern of the polarization
analog of the Hong-Ou-Mandel interferometer is replaced
by more complex patterns which may be composed, e.g.,
of two or three dips.

Properties of two (partially) overlapping entangled two-photon fields have
been also investigated in detail. It has been
shown that two entangled two-photon fields are mutually coherent
if the corresponding pump pulses are coherent. Partial
overlap of two entangled two-photon pulsed fields leads to higher
values of the visibility of coincidence-count interference pattern
in comparison with those appropriate
for one entangled two-photon field. This may be conveniently used
in many multiparticle experiments for which high values
of visibilities are required.

\begin{acknowledgement}
The author thanks for a kind hospitality in Quantum Imaging
Laboratory at Boston University. He also thanks J. Pe\v{r}ina
for reading the manuscript. He acknowledges support from
Grant No. VS96028 of the Czech Ministry of Education and
Grant No. 19982003012 of the Czech Home Department.
\end{acknowledgement}

\appendix
\section{Coincidence-count interference pattern for two Gaussian
pump pulses}

The quantities $ \rho_1 $ in Eq. (24), $ \rho_2 $ in Eq. (25),
$ R_{01} $ in Eq. (27), and $ R_{02} $ in Eq. (28) can be
expressed in the following forms when Gaussian pump pulses are
considered:
\begin{eqnarray}   
  \rho_1(l) &=& \frac{ \sqrt{\pi}|C_{\cal A}|^2 }{ 2\sqrt{2} R_0 D^2 }
  \int_{-DL/2 +|\Delta\tau_l|}^{DL/2 -|\Delta\tau_l|} d\tau
  \nonumber \\
  & & \mbox{} \hskip-6mm \times
   \left\{ \sum_{j=1,2}
   \xi_j^2 \tau_j \exp\left[ - 2\left(\frac{|\alpha_j|\tau_j
   \Lambda \tau}{ D} \right)^2 \right] \right\}, \\
  \rho_2(l,\vartheta,\phi) &=& \frac{ \sqrt{\pi} |C_{\cal A}|^2
  \xi_1 \xi_2}{ R_0 D^2 }
  \int_{-DL/2 +|\Delta\tau_l|}^{DL/2 -|\Delta\tau_l|} d\tau
  \nonumber \\
  & & \mbox{} \times
  {\rm Re} \left\{ \frac{ \exp(-i\phi) }{\sqrt{\alpha_1 + \alpha_2^*} }
   \right. \nonumber \\
  & & \left. \mbox{} \times \exp \left[ -
  \frac{\alpha_1\alpha_2^*}{\alpha_1 + \alpha_2^*}
  \left( \vartheta - \frac{2\Lambda}{D} \tau \right)^2 \right]
   \right\} , \\
  R_{01} &=& \frac{ \sqrt{\pi}|C_{\cal A}|^2 L }{ 2\sqrt{2}|D| }
   \left[ \xi_1^2 \tau_1 + \xi_2^2 \tau_2 \right] , \\
  R_{02}(\vartheta,\phi) &=& \frac{ \sqrt{\pi} |C_{\cal A}|^2 L \xi_1
   \xi_2}{ |D| } {\rm Re} \left\{ \frac{ \exp(-i\phi) }{\sqrt{
   \alpha_1 + \alpha_2^*} } \right. \nonumber \\
  & & \left. \mbox{} \times  \exp\left( -
   \frac{\alpha_1 \alpha_2^*}{\alpha_1 + \alpha_2^*}  \vartheta^2
   \right) \right\} ,
\end{eqnarray}
where $ \alpha_j = (1 + ia_j)/\tau_j^2 $ for $ j=1,2 $.

\end{document}